\documentstyle[12pt,epsf]{article}
\catcode`\@=11
\long\def\@makefntext#1{
\protect\noindent \hbox to 3.2pt {\hskip-.9pt  
$^{{\ninerm\@thefnmark}}$\hfil}#1\hfill}		%CAN BE USED 

\def\@makefnmark{\hbox to 0pt{$^{\@thefnmark}$\hss}}  %ORIGINAL 
	
\def\ps@myheadings{\let\@mkboth\@gobbletwo
\def\@oddhead{\hbox{}
\rightmark\hfil\ninerm\thepage}   
\def\@oddfoot{}\def\@evenhead{\ninerm\thepage\hfil
\leftmark\hbox{}}\def\@evenfoot{}
\def\sectionmark##1{}\def\subsectionmark##1{}}

\newcounter{sectionc}\newcounter{subsectionc}\newcounter{subsubsectionc}
\renewcommand{\section}[1] {\vspace*{0.6cm}\addtocounter{sectionc}{1} 
\setcounter{subsectionc}{0}\setcounter{subsubsectionc}{0}\noindent 
	{\normalsize\bf\thesectionc. #1}\par\vspace*{0.4cm}}
\renewcommand{\subsection}[1] {\vspace*{0.6cm}\addtocounter{subsectionc}{1} 
	\setcounter{subsubsectionc}{0}\noindent 
	{\normalsize\it\thesectionc.\thesubsectionc. #1}\par\vspace*{0.4cm}}
\renewcommand{\subsubsection}[1]
{\vspace*{0.6cm}\addtocounter{subsubsectionc}{1}
	\noindent {\normalsize\rm\thesectionc.\thesubsectionc.\thesubsubsectionc. 
	#1}\par\vspace*{0.4cm}}

\newcounter{appendixc}
\newcounter{subappendixc}[appendixc]
\newcounter{subsubappendixc}[subappendixc]

\renewcommand{\appendix}[1] {\vspace*{0.6cm}
        \refstepcounter{appendixc}
        \setcounter{figure}{0}
        \setcounter{table}{0}
        \setcounter{equation}{0}
        \renewcommand{\thefigure}{\Alph{appendixc}.\arabic{figure}}
        \renewcommand{\thetable}{\Alph{appendixc}.\arabic{table}}
        \renewcommand{\theappendixc}{\Alph{appendixc}}
        \renewcommand{\theequation}{\Alph{appendixc}.\arabic{equation}}
        \noindent{\bf Appendix \theappendixc #1}\par\vspace*{0.4cm}}

\def\abstracts#1{{
	\centering{\begin{minipage}{12.2truecm}\normalsize\baselineskip=12pt\noindent
	\centerline{\footnotesize ABSTRACT}\vspace*{0.3cm}
	\parindent=0pt #1
	\end{minipage}}\par}} 

\newcounter{itemlistc}
\newcounter{romanlistc}
\newcounter{alphlistc}
\newcounter{arabiclistc}

\newcommand{\fcaption}[1]{
        \refstepcounter{figure}
        \setbox\@tempboxa = \hbox{\footnotesize Fig.~\thefigure. #1}
        \ifdim \wd\@tempboxa > 6in
           {\begin{center}
        \parbox{6in}{\footnotesize\baselineskip=12pt Fig.~\thefigure. #1}
            \end{center}}
        \else
             {\begin{center}
             {\footnotesize Fig.~\thefigure. #1}
              \end{center}}
        \fi}

\newcommand{\tcaption}[1]{
        \refstepcounter{table}
        \setbox\@tempboxa = \hbox{\footnotesize Table~\thetable. #1}
        \ifdim \wd\@tempboxa > 6in
           {\begin{center}
        \parbox{6in}{\footnotesize\baselineskip=12pt Table~\thetable. #1}
            \end{center}}
        \else
             {\begin{center}
             {\footnotesize Table~\thetable. #1}
              \end{center}}
        \fi}

\def\@citex[#1]#2{\if@filesw\immediate\write\@auxout
	{\string\citation{#2}}\fi
\def\@citea{}\@cite{\@for\@citeb:=#2\do
	{\@citea\def\@citea{,}\@ifundefined
	{b@\@citeb}{{\bf ?}\@warning
	{Citation `\@citeb' on page \thepage \space undefined}}
	{\csname b@\@citeb\endcsname}}}{#1}}

\newif\if@cghi
\def\cite{\@cghitrue\@ifnextchar [{\@tempswatrue
	\@citex}{\@tempswafalse\@citex[]}}
\def\citelow{\@cghifalse\@ifnextchar [{\@tempswatrue
	\@citex}{\@tempswafalse\@citex[]}}
\def\@cite#1#2{{$\null^{#1}$\if@tempswa\typeout
	{IJCGA warning: optional citation argument 
	ignored: `#2'} \fi}}

 1
 1
 1

\font\ninerm=cmr9

\begin{document}
\pagestyle{empty}
\hfill{ANL-HEP-CP-96-33}
\vskip 0.2cm
\hfill{May 1, 1996}
\vskip 3cm
\centerline{\normalsize\bf ASSOCIATED PRODUCTION OF CHARM AND A HARD
PHOTON} 
\vskip 1.cm
%\begin{center}
\centerline{B. Bailey$^a$, E. L. Berger$^b$ and L. E.
Gordon$^b$}
\vskip 0.3in
\centerline{\it $^a$Physics Department, 
Eckerd College, St. Petersburg, FL 33711, U.S.A.}
\vskip 0.2in
\centerline{\it $^b$High Energy Physics Division, Argonne National Laboratory, 
Argonne, IL 60439, U.S.A.}
%\end{center}
\vskip 2cm
%\centerline{April 1996}
%\vskip .5cm
%\centerline{\bf Abstract}
\abstracts{ 
The two particle inclusive cross section for the reaction $p 
+\bar{p}\rightarrow \gamma + c + X$ is studied in perturbative quantum
chromodynamics at order $O(\alpha ^2_s)$, for large values of the transverse 
momentum of the prompt photon and charm quark. Two different techniques
are used in performing the phase-space integrals; the first is purely
analytical, and the second is a combination of analytic and Monte Carlo 
integration methods. The second, more versatile technique facilitates 
imposition of photon isolation restrictions and other
selections of relevance in experiments.  Differential distributions are 
provided for various observables, and a comparison is made with
preliminary data from the CDF collaboration.}
\vskip 6cm
Invited talk presented by L. E. Gordon at the XXXI st Rencontres de
Moriond, `QCD and High Energy Hadronic Interactions' March 23-30, 1996,
Les Arcs, France

\newpage

\section{Introduction}
\pagestyle{plain}
The observation of a photon carring a large transverse momentum, $p_T$, among 
the final state products of a high energy hadronic reaction has long been 
regarded as a powerful tool for gathering information about the short distance
dynamics of these reactions, primarily because photons couple directly
to quarks via the point-like electromagnetic interaction. Single prompt
photon production has been studied extensively, both theoretically
and experimentally, and has provided useful information about hadronic
structure, in particular the gluon structure function of the proton.  Data
are beginning to appear from the study of the production
of a prompt photon ($\gamma$) in association with a heavy quark whose
transverse momentum balances a substantial portion of that of the
photon. This two-particle inclusive 
reaction is particularly interesting because it offers
the possiblility of a detailed study of the underlying QCD dynamics such
as, e.g., rapidity correlations. In addition, it may provide a
direct measurement of the charm content of the proton.

We report here on two next-to-leading order perturbative QCD calculations 
we have done of the reaction $p +\bar{p}\rightarrow \gamma + 
c + X$ at high energy $^{1,2)}$. In these calculations two different
techniques are used in performing the phase-space integrals. In the first, 
purely analytical techniques are used.  In the second approach, we use a
combination of analytical and Monte Carlo techniques, which is more
flexible and allows implementation of isolation cuts and other experimentally 
relevant selections. To warrant use of 
perturbation theory and the massless approximation, we
limit our considerations to values of transverse momenta of the photon
and charm quark $p^{\gamma,c}_{T} > 10$ GeV.  

Only one {\it {direct}} hard scattering subprocess contributes in 
leading order: the quark-gluon Compton subprocess $g c \rightarrow \gamma c$.
The initial charm quark and the initial gluon are constituents of the initial 
hadrons.  In addition, there is a leading order {\it {fragmentation}} process 
in which the photon is produced from quark or gluon fragmentation, e.g.,
$g g \rightarrow c \bar{c}$ followed by $\bar{c} \rightarrow \gamma X$, or
$q c \rightarrow  q c$ followed by $q \rightarrow \gamma$.
At next-to-leading order in QCD, several subprocesses contribute to the
$\gamma + c +X$ final state: $gc \rightarrow gc\gamma$,
$g g \rightarrow c \bar{c} \gamma$,
$q \bar{q} \rightarrow c \bar{c} \gamma$,
$q c \rightarrow  q c \gamma$,
$\bar{q} c \rightarrow \bar{q} c \gamma$,
$c \bar{c} \rightarrow c \bar{c} \gamma$, and
$c c \rightarrow c c \gamma$.
A full next-to-leading order calculation requires the computation of the 
hard-scattering matrix elements for these two-to-three particle production 
processes as well as the one-loop $O(\alpha_s^2)$ corrections to the lowest 
order subprocess $g c \rightarrow \gamma c$.     
At $O(\alpha ^2_s)$ there are, in addition, fragmentation processes
in which the hard-scattering two-particle final-state subprocesses
$c+g \rightarrow \gamma+ c, c+\bar{c}\rightarrow \gamma+g$
and $q+\bar{q}\rightarrow \gamma +g$
are followed by fragmentation processes $c\rightarrow c X$, in the case of
the first subprocess, and
$g\rightarrow c X$ in the cases of the last two. These should be included
because we factor the collinear singularities in the
corresponding three-body final-state processes into non-perturbative
fragmentation functions for production of a charm quark from a particular
parton.

We are interested in the two-particle inclusive differential cross section, 
$E_\gamma E_cd\sigma/d^3p_\gamma d^3p_c$, where
$(E,p)$ represents the four-vector momentum of the $\gamma$ or $c$ quark.  For 
each contributing subprocess, this calculation requires integration over the
momentum of the unobserved final parton in the two-to-three particle 
subprocesses
($g$, $\bar{c},q$, or $\bar{q}$).  Collinear singularities must be handled
analytically by dimensional regularization and absorbed into parton momentum
densities or fragmentation functions. 

\section{Analytic Calculation}

In our purely analytical calculation$^{1}$, we must integrate over enough of
the phase-space analytically in order to cancel all soft and collinear 
singularities. 
This procedure imposes technical limitations such that it is not possible to 
obtain the fully differential two-particle cross section.  
We calculate instead the cross section
$d\sigma/dp_T^\gamma dy^\gamma dz$, where the variable $z$ is defined by
\begin{equation}
z=-\frac{p^\gamma_T.p^c_T}{(p^\gamma_T)^2}.
\end{equation}
This definition indicates that $z$ contains information on the
transverse momentum of the charm quark, but no information on its
rapidity $y^c$. Moreover, it is not possible to implement isolation cuts
on the photon in the purely analytic calculation. Nevertheless the calculation 
is valuable for obtaining the relative importance of the
various subprocess contributions and the overall the size of the cross section, 
as well as for checking aspects of the more versatile Monte Carlo calculation.

We calculate the cross section
$d\sigma/dp_T^\gamma dy^\gamma dz$.  Whenever 
the charm
quark and photon have balancing values of $p_T$ then $z=1$.  This occurs
\begin{enumerate}
\item{for the leading order process $cg\rightarrow \gamma c$,}
\item{for the virtual gluon exchange processes,}
\item{whenever a gluon becomes soft in the three-body processes, or}
\item{when a third parton becomes collinear to the beam in a three-body
process.}
\end{enumerate}
The point $z=1$ is associated with various soft and collinear poles,
and in order to expose these poles all phase-space integrals are
carried out in $4-2\epsilon$ dimensions.  The poles are exposed
by expanding the results in $(1-z)$. The analytical cross section
is expressed in terms of distributions in $(1-z)$ such as $1/(1-z)_+$
and $\delta(1-z)$. In addition whenever the final state photon becomes
collinear to a final state quark, the resulting singularity must be
exposed in a similar way before it is absorbed into the photon fragmentation 
function. This singularity occurs at another value of $z$, ($z=z_1$).
This means that there are also distributions in $(z-z_1)$. The cross
section is singular at $z=1$ and $z=z_1$, but the singularities are
integrable through the use of `plus'-distributions. The cross section
is presented in terms of integrals over finite regions of $z$
such as
\begin{eqnarray}
\frac{d\sigma}{dp_T^\gamma dy^\gamma dz}&=&\frac{1}{\Delta z}\int^{z+
\frac{\Delta
z}{2}}_{z-\frac{\Delta z}{2}}\frac{d\sigma}{dp_T^\gamma dy^\gamma
dz'}dz' ;\nonumber \\
\frac{d\sigma}{dp_T^\gamma dy^\gamma}&=&\int^{z_b}_{z_a}
\frac{d\sigma}{dp_T^\gamma dy^\gamma
dz'}dz'.
\end{eqnarray}

Our results are presented at a center-of-mass energy $\sqrt{s}=1.8$ TeV.
The analytical results are obtained with the GRV parametrization of the proton
parton densities, and all renormalization/factorization scales are
taken as $\mu=p_T^\gamma$. Up to $30\%$ differences are observed if the
CTEQ3M parton distributions are used. We sum over charm and anticharm
production throughout. 
\begin{figure}
%\vspace*{18pt}
%\leftline{\hfill\vbox{\hrule width 5cm height0.001pt}\hfill}
{\hskip 0.3cm}\hbox{\epsfxsize7.5cm\epsffile{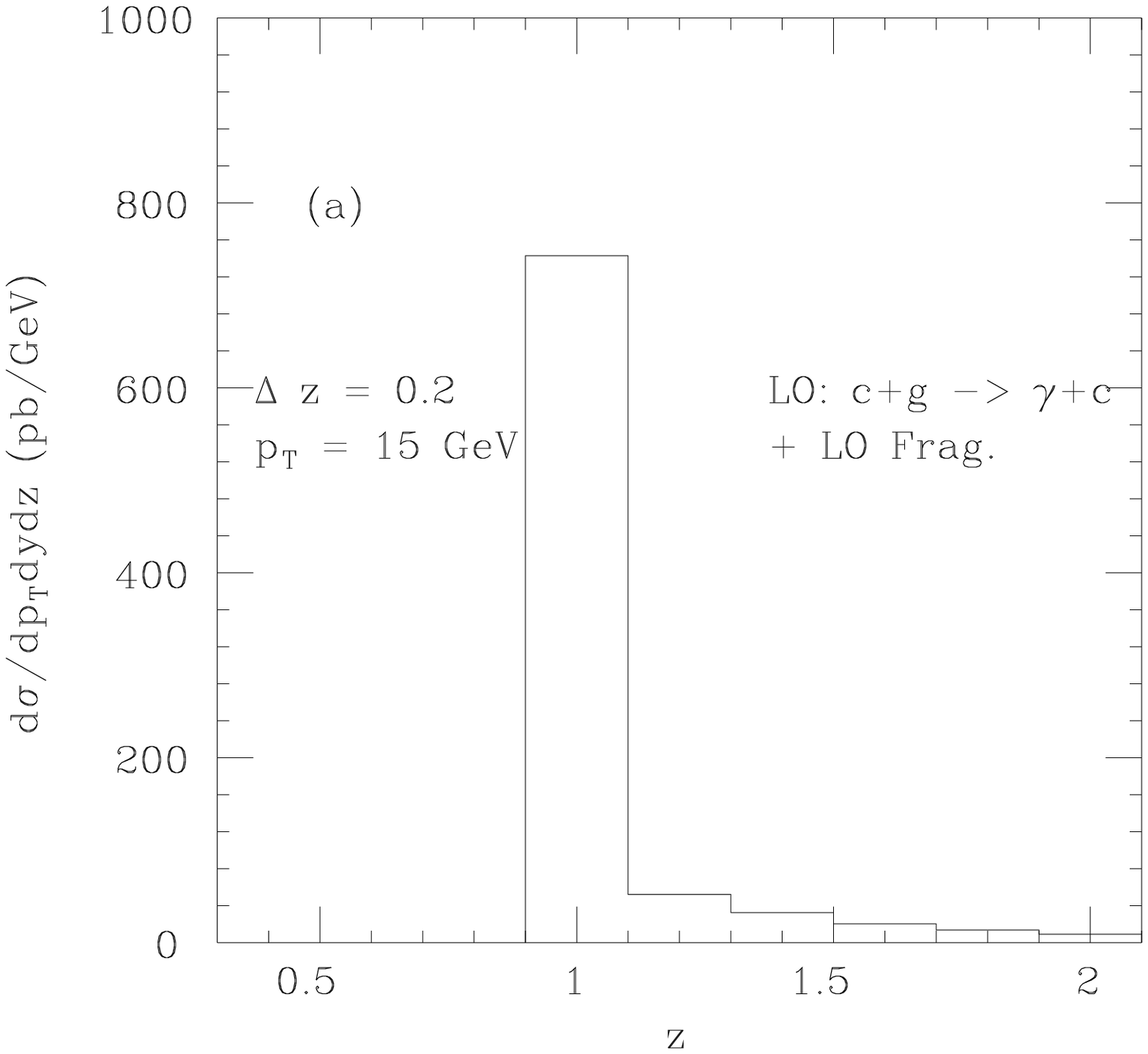}{\hskip 0.3cm}
\epsfxsize7.5cm\epsffile{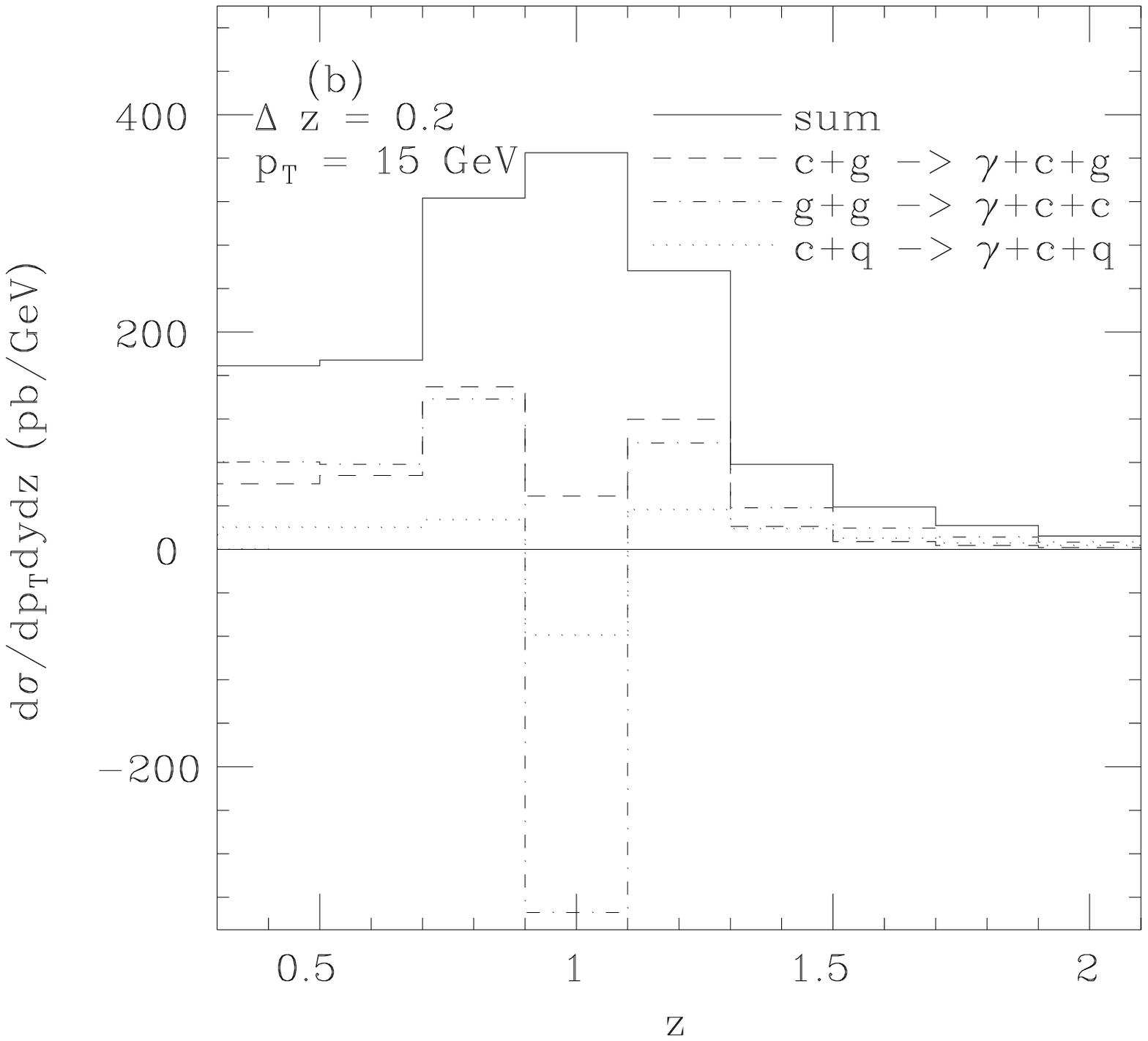}}
%\vspace*{1.4truein}             %ORIGINAL SIZE=1.6TRUEIN x 100% - 0.2TRUEIN
%\leftline{\hfill\vbox{\hrule width 5cm height0.001pt}\hfill}
\fcaption{Cross section as a function of z at $y^\gamma=0$ in (a)
leading order and (b) next-to-leading showing the dominant
contributions.}
%\label{fig:fone}
\end{figure}
In Fig.1a the net lowest order contribution is shown as a function of $z$
at $p_T^\gamma=15$ GeV and for a bin size $\Delta z=0.2$. The lowest
order cross section is made up of the lowest order direct term
$cg\rightarrow\gamma c$, which is proportional to $\delta(1-z)$ and
provides the peak at $z=1$, plus the various photon fragmentation
contributions which contribute in the region $z\geq 1$. The most
significant feature of this curve is that there is no contribution to
the cross section in the region $z\leq 1$. This 
unrealistic prediction shows the inadequacy of the lowest order
predictions.

Figure 1b shows the distribution in $z$ predicted by the next-to-leading 
order calculation. The next-to-leading order contributions serve to lower the 
peak at $z=1$, and they broaden the distribution. The cross section is finite 
at all values of $z$, closer to the
situation observed in experiments. In addition, in Fig.1b we
display contributions from the most important subprocesses. 
The $cg$ initiated process dominates the cross section, but there are important
contributions from the $gg$ and $cq$ initiated process in the low $p_T^\gamma$ 
region. 

\section{Monte Carlo Calculation}

The combination of analytic and Monte Carlo techniques used here to
perform the phase space integrals is documented and described elsewhere$^2$. 
The analytic/Monte Carlo method 
allows for the calculation of many different observables without 
much extra effort and for the imposition of experimental cuts.
In collider experiments a photon is observed and its momentum is well
measured only when the photon is isolated from neighboring hadrons.  In
our calculation, we impose isolation in terms of the cone variable $R$:  
\begin{equation}
\sqrt{(\Delta y)^2 + (\Delta \phi)^2} \leq R.     \label{eq:Rdef}
\end{equation}
In Eq.~(\ref{eq:Rdef}), $\Delta y$ ($\Delta \phi$) is the difference between 
the rapidity (azimuthal angle in the transverse plane) 
of the photon and that of any parton in the final state.  The photon is said
to be isolated in a cone of size $R$ if the ratio of the hadronic energy in the 
cone and the transverse momentum of the photon does not exceed 
$\epsilon = {\rm 2 GeV}/p_T^{\gamma}$. We show distributions
for the choice $R = 0.7$  typical of current experiments.  

The structure of the QCD hard-scattering matrix element produces 
{\it {positive}} correlations in rapidity at collider energies. 
To examine these correlations more precisely, 
in Fig. 2a, we display the differential cross section in $y^c$, for 
two intervals of $y^{\gamma}$ in the forward rapidity region.  These
distributions show that the typical rapidity of the charm quark follows that
of the photon, and thereby confirms the positive correlation
between the rapidities of the photon and charm quark. 
\begin{figure}
\vspace*{18pt}
%\leftline{\hfill\vbox{\hrule width 5cm height0.001pt}\hfill}
{\hskip 0.3cm}\hbox{\epsfxsize7.7cm\epsffile{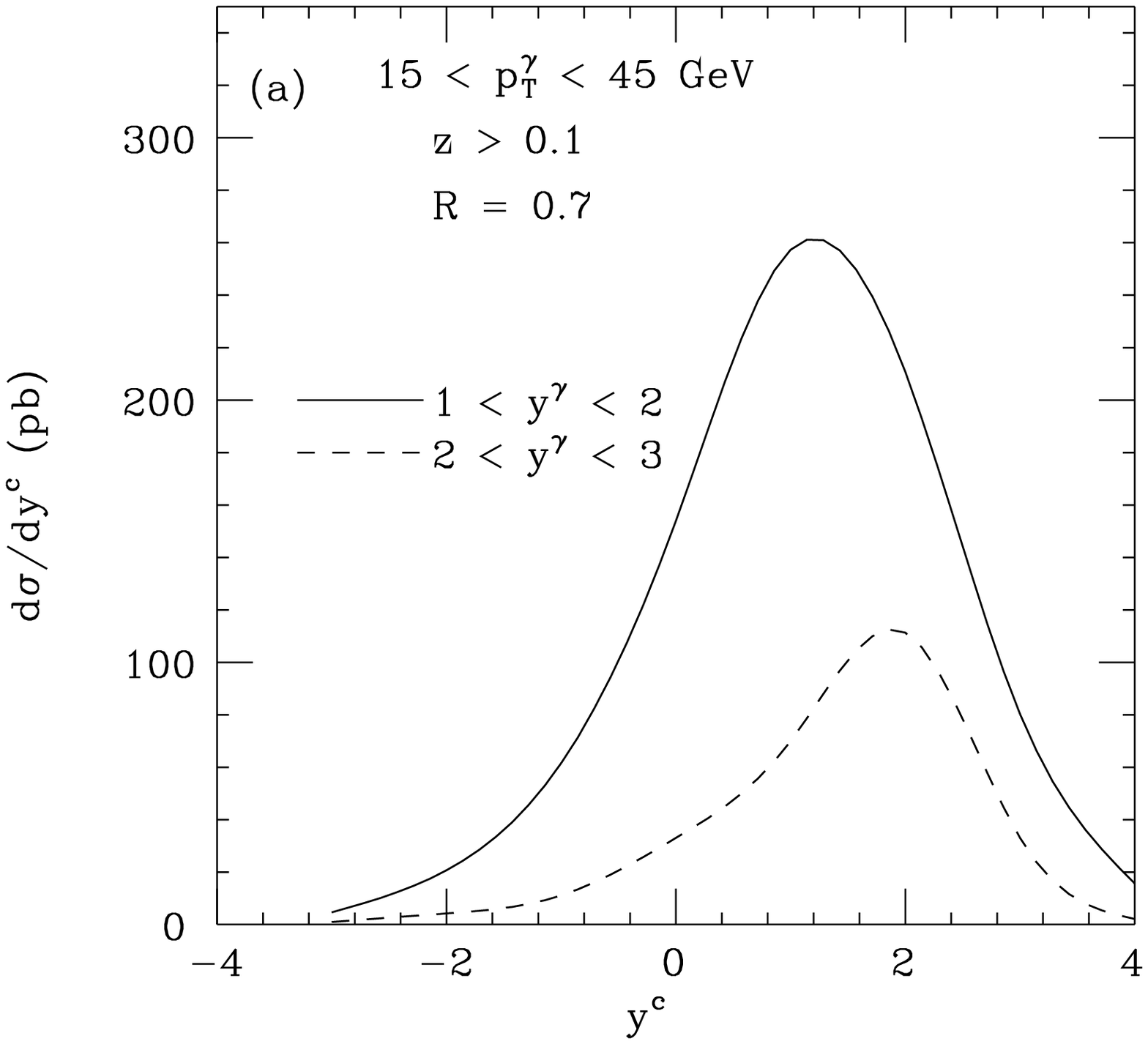}{\hskip 0.3cm}
\epsfxsize7.7cm\epsffile{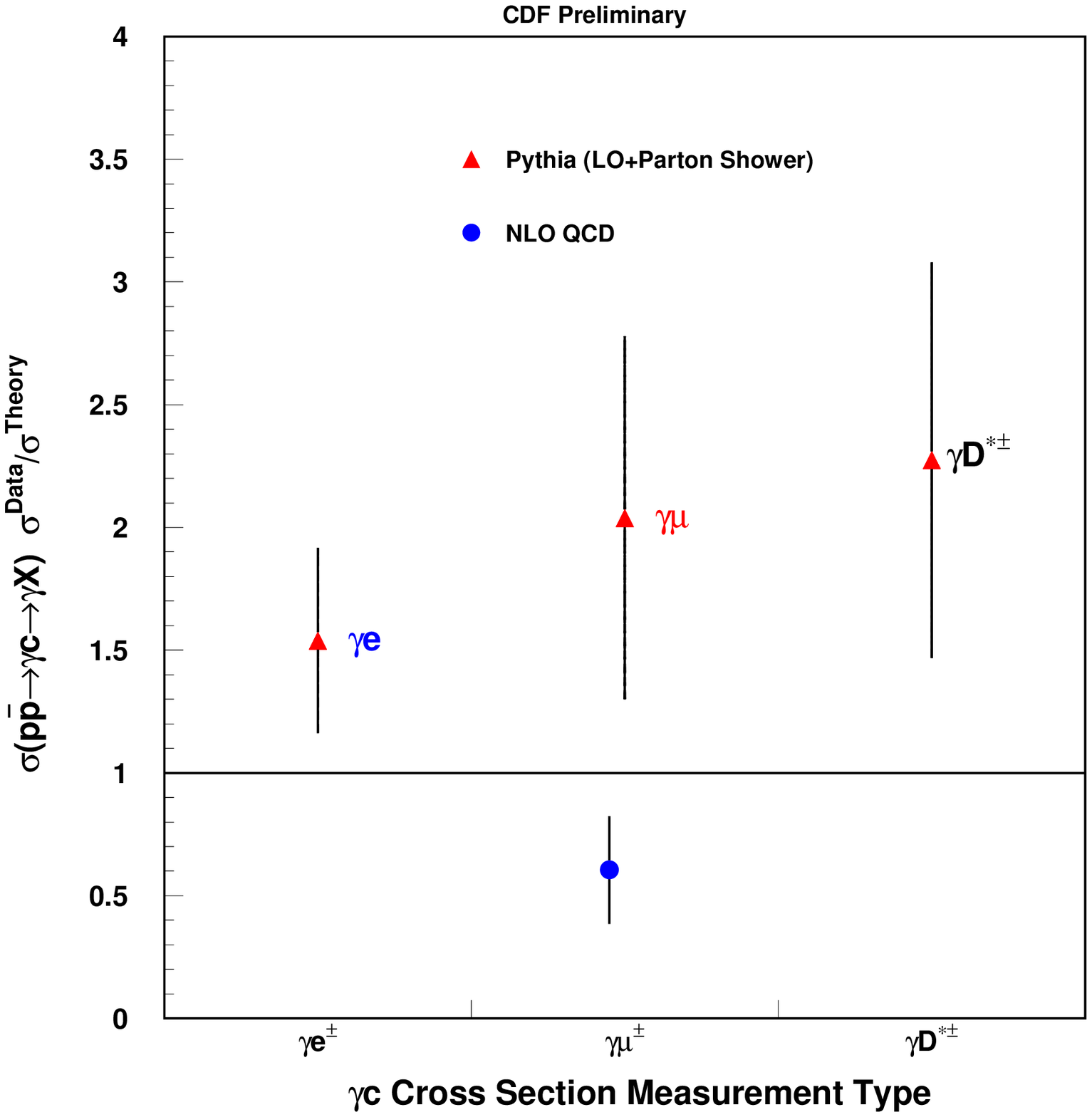}}
%\vspace*{1.4truein}             %ORIGINAL SIZE=1.6TRUEIN x 100% - 0.2TRUEIN
%\leftline{\hfill\vbox{\hrule width 5cm height0.001pt}\hfill}
\fcaption{ (a)Cross section $d\sigma/dy^c$ as a function of the rapidity of 
the charmquark for $p +\bar{p}\rightarrow \gamma + c + X$ at $\sqrt{s}=1.8$ 
TeV. The solid curve shows the result when the photon rapidity is restricted to 
$1.0 < y^\gamma < 2.0$, and the dashed curve displays the result for 
$2.0 < y^\gamma < 3.0$.
(b) Ratio of the measured cross section to that predicted by theory for
various final state charm decay products.}
%\label{fig:fone}
\end{figure}
In Fig.2b we compare our results to CDF data$^3$ for
photon plus $\mu^\pm$ production.  The three upper points are
obtained$^4$ from the Monte Carlo event generator Pythia whereas the lower
point is that given by our theoretical calculation. The Pythia results
lie substantially below the data whereas our results are higher than the
data but somewhat closer to it.  

\section{Conclusions}

We presented the results of two calculations of
the inclusive production of a prompt photon in association with a heavy quark at
large values of transverse momentum.  Both analyses are done at
next-to-leading order in perturbative QCD.  
Our results agree quantitatively as they should, but the 
combination of analytic and Monte Carlo methods is more
versatile.  We provide differential cross sections in transverse
momenta and rapidity, including
photon isolation restrictions, that should facilitate contact with experimental
results at hadron collider energies.  We show that the study of
two-particle inclusive distributions, with specification
of the momentum variables of both the final prompt photon and the final heavy
quark, tests correlations inherent in the QCD matrix elements 
and should provide a means for measuring the charm quark density in the 
nucleon.  A comparison of our next-to-leading predictions with the preliminary 
CDF data shows reasonable agreement. 

The work at Argonne National Laboratory was supported by the US Department of
Energy, Division of High Energy Physics, Contract number W-31-109-ENG-38.
This work was supported in part by Eckerd College.

\section{References}
\newcounter{num}
\begin{list}%
{[\arabic{num}]}{\usecounter{num}
    \setlength{\rightmargin}{\leftmargin}}
\item {E. L. Berger and L. E. Gordon, Argonne report 
ANL-HEP-PR-95-36 (hep-ph /9512343), Phys. Rev. {\bf D} (in press),
and references therein.}
\item {B. Bailey, E. L. Berger and L. E. Gordon, Argonne report 
ANL-HEP-PR-95-87 (hep-ph/9602373), submitted to Phys. Rev. {\bf D},
and references therein.}
\item{CDF Collaboration, R. Blair {\it{et al}},
Proceedings of the 10th Topical Workshop on Proton-Antiproton Collider
Physics, May, 1995 (AIP Conference Proceedings 357), edited by R. Raja and
J. Yoh (AIP Press, N.Y., 1996), pp 557-567.}
\item{S. Kuhlmann, CDF Collaboration, private communication.}
\end{list}

\end{document}